\documentclass[12pt]{iopart}
\usepackage{graphics}
\begin{document}
\jl{1}
\title
[Magnetization plateau in the $S=1/2$ trimerized $XXZ$ chain]
{Magnetization plateau and quantum phase transition
of the $S=1/2$ trimerized $XXZ$ spin chain}
\author{Kiyomi Okamoto$\dagger$ and Atsuhiro Kitazawa$\ddagger$}
\address{$\dagger$Department of Physics,
         Tokyo Institute of Technology,
         Oh-Okayama, Meguro-ku, Tokyo 152-8551, Japan}
\address{$\ddagger$Department of Physics, Kyushu University 33,
         Fukuoka, 812-8581, Japan}
\begin{abstract}
We study the plateau of the magnetization curve at $M = M_{\rm s}/3$
($M_{\rm s}$ is the saturation magnetization)
of the $S=1/2$ trimerized $XXZ$ spin chain.
By examining the level crossing of low-lying excitations
obtained from the numerical diagonalization,
we precisely determine the phase boundary between the plateau state
and the no-plateau state on the $\Delta-t$ plane,
where $\Delta$ denotes the $XXZ$ anisotropy and $t$ the magnitude
of the trimerization.
This quantum phase transition is of the Berezinskii-Kosterlitz-Thouless
type.
\end{abstract}
%
\maketitle
\section{Introduction}
In recent years the quantized plateau of the magnetization curve
of spin chains has been attracting much attention.
Hida~\cite{Hida} numerically studied the $S=1/2$
ferromagnetic-ferromagnetic-antiferromagnetic trimerized Heisenberg chain
and found the plateau of the magnetization curve at $M=M_{\rm s}/3$
($M_{\rm s}$ is the saturation magnetization)
for some parameter region of $J_{\rm F}/J_{\rm AF}$,
where $J_{\rm F}$ and $J_{\rm AF}$ are the ferromagnetic and
antiferromagnetic couplings, respectively.
One of the present authors (K.O.)~\cite{Okamoto-F-F-AF} analytically
investigated Hida's model to clarify the mechanism for the
appearance and disappearance of the $M=M_{\rm s}/3$ plateau.
Later related numerical and theoretical~\cite{Roji-Miyashita,
Tonegawa-Nakao-Kaburagi,Tonegawa-Nishida-Kaburagi,Totsuka, OYA,
Sakai-Takahashi} are reported in the literature.
The magnetization plateaus are also found experimentally
in $S=1$ Ni compound
${\rm [Ni_2 (Medpt)_2 (\mu\mbox{-}ox)(\mu\mbox{-}N_3)]ClO_4 \cdot 0.5 H_2 O}$
~\cite{Narumi}
and in $S=1/2$ Cu compound ${\rm NH_4 Cu Cl_3}$~\cite{Shiramura}.
The behaviour of the magnetization curve of ${\rm NH_4 Cu Cl_3}$
is quite remarkable, because magnetization plateaus observed at
$M = (3/4) M_{\rm s}$ and $M = (1/4)M_{\rm s}$
but not at $M = 0$ and $M = (1/2) M_{\rm s}$.

Oshikawa, Yamanaka and Affleck~\cite{OYA} gave the necessary condition for the
appearance of the magnetization plateau
\begin{equation}
  n (S - \langle m \rangle) = {\rm integer}
  \label{eq:OYA}
\end{equation}
where $n$ is the periodicity of the state, $S$ the magnitude of spins
and $\langle m \rangle$ the average magnetization per one spin.
Since (\ref{eq:OYA}) is the necessary condition,
it depends on the details of the models whether the magnetization plateau
exists or not, even if the condition (\ref{eq:OYA}) is satisfied.

\begin{figure}[h]
   \begin{center}
      \scalebox{0.5}[0.5]{\includegraphics{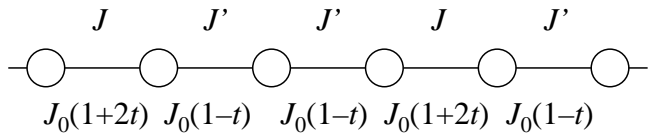}}
   \end{center}
   \caption{Sketch of the trimerized $XXZ$ chain.
            The expression of the lower line corresponds
             to the parametrized form~(\ref{eq:ham2}).}
   \label{fig:model}
\end{figure}

In this paper we study the $M=M_{\rm s}/3$ plateau of the
$S=1/2$ trimerized $XXZ$ spin chain described by
\begin{equation}
  H = \sum_{j=1}^{L}\left\{
  J^{'}\left[ h_{3j-2,3j-1}(\Delta)+h_{3j-2,3j-1}(\Delta) \right]
  +J h_{3j,3j+1}(\Delta)\right\}
\label{eq:ham1}
\end{equation}
where
\begin{equation}
 h_{l,m}(\Delta) = S^{x}_{l}S^{x}_{m}+S^{y}_{l}S^{y}_{m}
    +\Delta S^{z}_{l}S^{z}_{m}\,.
\end{equation}
Our model is sketched in figure~\ref{fig:model}.

In \S2 we qualitatively discuss the properties of the transition
between the plateau and no-plateau states by use of the bosonized Hamiltonian.
In \S3 we determine the phase boundary of between the plateau and no-plateau
states from the numerical diagonalization data by examining the
crossings of the low-lying excitations~\cite{Nomura-Kitazawa}.
\S4 is devoted to discussion.
\section{Transition between the plateau state and the no-plateau state}
It is convenient to parametrize the Hamiltonian (\ref{eq:ham1}) as
\begin{eqnarray}
  H &=& J_{0}\sum_{j=1}^{L}\left\{
  (1-t)[ h_{3j-2,3j-1}(\Delta)+h_{3j-2,3j-1}(\Delta)]\right. \nonumber \\
  && \left. +(1+2t)h_{3j,3j+1}(\Delta)\right\},
\label{eq:ham2}
\end{eqnarray}
where
\begin{equation}
  J_{0}=\frac{2J^{'}+J}{3}~~~~~~~
  t = - \frac{J^{'}-J}{2J^{'}+J}\,.
\end{equation}
The model is sketched in figure~\ref{fig:model}.
The bosonized expression of the Hamiltonian (\ref{eq:ham2}) can be obtained
by the following procedure:

\noindent
\leftskip5truemm
(a) Transforming (\ref{eq:ham2}) into the spinless fermion expression
by use of the Jordan-Wigner transformation.
The spacing between the neighboring spins is taken as the unit length.

\noindent
(b) Linearizing the dispersion relation of the spinless fermions
$\omega(k) = J_0 \cos k$ around $k=\pm k_{\rm F}$,
where $k_{\rm F} \equiv \pi/3$ corresponds to the band filling
of $M = M_{\rm s}/3$.
The Fermi velocity at $k = k_{\rm F}$ is $v_{\rm F} = (\sqrt{3}/2)J_0$.

\noindent
(c) Taking the effects of trimerization and the interactions between fermions
into account through the procedure
similar to that of the standard bosonization technic.

\leftskip0truecm
\noindent
From the above procedure, we obtain the following sine-Gordon Hamiltonian
\begin{equation}
  H
  = \frac{1}{2\pi} \int \rmd x\left[ v_{\rm s}K(\pi\Pi)^{2}
    + \frac{v_{\rm s}}{K}
   \left(\frac{\partial \phi}{\partial x}\right)^{2}\right]
   + \frac{y_{\phi}v_{\rm s}}{2\pi}\int \rmd x\cos\sqrt{2}\phi
\label{eq:sg}
\end{equation}
where $v_{\rm s}$ is the spin wave velocity of the system,
$\Pi$ is the momentum density conjugate to $\phi$,
$[\phi(x),\Pi(x')] = i\delta(x-x')$, and
the coefficients $v_{\rm s}$, $K$, and $y_{\phi}$ are related to
$J_{0}$, $t$ and $\Delta$ as
\begin{equation}
  v_{\rm s} = \sqrt{3}J_{0} \sqrt{AC}~~~~~
  K = \frac{1}{2\pi}\sqrt{\frac{C}{A}}~~~~~
  y_{\phi}v_{\rm s} = 2\pi J_{0}t
  \label{eq:K}
\end{equation}
where
\begin{equation}
  A = \frac{1}{8\pi}\left( 1+\frac{5}{\sqrt{3}\pi}\Delta\right)~~~~~
  C = 2\pi\left( 1 - \frac{1}{\sqrt{3}\pi}\Delta\right)\,.
  \label{eq:A-C}
\end{equation}
The dual field $\theta$ is defined by $\partial_{x}\theta = \pi\Pi$,
and we make the identification $\phi \equiv \phi+\sqrt{2}\pi$,
$\theta \equiv \theta+\sqrt{2}\pi$.
We note that the umklapp term (which exists in $M=0$ case and
is important to describe the transition between the spin-fluid state
and the N\'eel state) does not exist, because $2k_{\rm F}$ is not equal
to the reciprocal lattice wave numbers.
The field $\phi$ is related to the fast varying (in space) part
of the spin density $S^z(x)$ in the comtinuum picture as
\begin{equation}
    S^z_{\rm fast}(x)
    = {1 \over3}
      \left\{ \cos \left(2k_{\rm F}x - {\pi \over 3}+ \sqrt{2}\phi \right)
              + {1 \over 2} \right\}
\end{equation}
which makes it clear the physical meaning of $\phi$.
We note that the slowly varying part of the spin density
is proportional to $\partial \phi / \partial x$.

\begin{figure}[h]
   \begin{center}
      \scalebox{0.4}[0.4]{\includegraphics{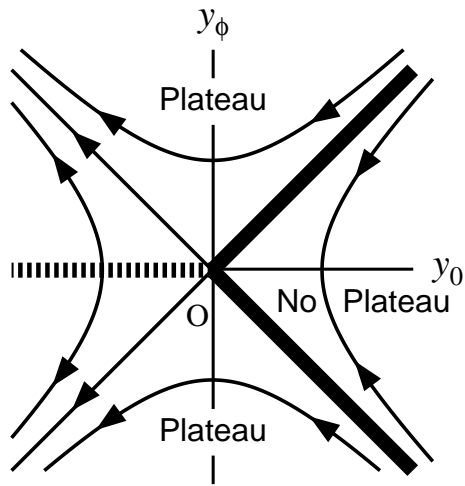}}
   \end{center}
   \caption{Flow diagram of the renormalization
            group equation~(\ref{eq:RG}).
            The thick solid lines show the BKT lines
            and the thick dotted line the Gaussian line.}
   \label{fig:RG-flow}
\end{figure}

As is well known, the excitation spectrum of the sine-Gordon model is
either massive or massless depending on the values of $K$ and $y_\phi$.
In the massive case the $M_{\rm s}/3$ magnetization plateau exists,
and in the massless case it does not~\cite{Okamoto-F-F-AF}.
It is convenient to discuss the properties of (\ref{eq:sg}) in the framework of
the renormalization group method.
The renormalization group equations for (\ref{eq:sg}) are
\begin{equation}
  \frac{\rmd K(L)^{-1}}{\rmd\ln L} = \frac{1}{8}y_{\phi}(L)^{2}~~~~~~~~~~
  \frac{\rmd y_{\phi}(L)}{\rmd\ln L} = \left( 2 - \frac{K(L)}{2}\right) y_{\phi}(L)
\end{equation}
where $L$ is an infrared cutoff.
Denoting $K(L)=4(1+y_{0}(L)/2)$ near $K(L)=4$, we obtain
\begin{equation}
  \frac{\rmd y_{0}(L)}{\rmd\ln L} = -y_{\phi}(L)^{2}~~~~~~~~~~
  \frac{\rmd y_{\phi}(L)}{\rmd\ln L} = -y_{0}(L)y_{\phi}(L)
  \label{eq:RG}
\end{equation}
and show its flow diagram in figure~\ref{fig:RG-flow}.
The Berezinskii-Kosterlitz-Thouless (BKT) transition
occurs at $y_{0}=|y_{\phi}|$, shown by thick solid lines.
At the BKT transition point, by substituting $y_0 = |y_\phi|$
into eqaution~(\ref{eq:RG}), we have
\begin{equation}
  y_{0}(L) = \frac{y_{0}}{y_{0}\ln(L/L_{0}) +1}
\label{eq:log}
\end{equation}
where $y_{0}$ is the bare value.
When $y_0<0$ ({\it i.e.}, $K<4$), any small (but not equal to zero) amount
of trimerization brings about the magnetization plateau.
The phase boundary between two plateau regions
is Gaussian line (thick dotted line), on which the critical exponents
vary continuously.
In the no-plateau region, the effect of the trimerization vanishes
in the sense of the renormalization group
due to the strong quantum fluctuations.

We note that it is dangerous to apply the conventional
phenomenological renormalization group method to the BKT transition,
as is fully discussed in~\cite{Okamoto-Nomura}.

\section{Numerical approach}
The scaling dimension of the primary field
${\cal{O}}_{m,n}= \exp(m\sqrt{2}\phi+n\sqrt{2}\theta)$ for $y_{\phi}=0$
is given by
\begin{equation}
  x_{n,m} = \frac{K}{2}m^{2} + \frac{1}{2K}n^{2}
\end{equation}
where $n$ and $m$ are integers with the periodic boundary condition (PBC).
According to the finite size scaling theory by Cardy~\cite{Cardy1,Cardy2},
the excitation energy of the finite size system at a critical point
is related to the scaling dimension as
\begin{equation}
  x_{m,n}(L) = \frac{L}{2\pi v_{\rm s}}(E_{m,n}(L)-E_{g}(L))
\end{equation}
where $E_{g}(L)$ is the ground state energy of $L$-spin system with PBC.
Near the BKT transition ($K\approx 4$), the excitation energy is written as
\begin{equation}
  \frac{L}{2\pi v_{\rm s}}\Delta E_{m,0}(L) = 2m^{2}  + y_{0}(L)m^{2}
\label{eq:dimm}
\end{equation}
\begin{equation}
  \frac{L}{2\pi v_{\rm s}}\Delta E_{0,n}(L)
  = \frac{1}{8}n^{2}
  - y_{0}(L)\frac{1}{16}n^{2}
\label{eq:dimn}
\end{equation}
for integer $m$, $n$.
Thus, considering equation~(\ref{eq:log}),
we have the logarithmic corrections for finite size spectrum.

To determine the BKT transition point, we use the method
developed by Nomura and Kitazawa~\cite{Nomura-Kitazawa},
in which the level crossings for some excitations are used.
With the twisted boundary condition (TBC) $S^{x,y}_{3L+1}=-S^{x,y}_{1}$,
$S^{z}_{3L+1}=S^{z}_{1}$, the integer $m$ in the operator ${\cal{O}}_{m,n}$
shifts to $m+1/2$ as ${\cal{O}}_{m,n}\rightarrow {\cal{O}}_{m+1/2,n}$.
For the scaling dimensions of the operators
$\sqrt{2}\cos (\phi/\sqrt{2})$ and $\sqrt{2}\sin (\phi/\sqrt{2})$
we have the following finite size corrections
\begin{equation}
   \begin{array}{l}
     x^{\rm c}_{1/2,0}(L)
       =  \displaystyle\frac{1}{2}
          + \displaystyle\frac{1}{4}y_{0}(L)
          + \displaystyle\frac{1}{2}y_{\phi}(L) \\ \\
     x^{\rm s}_{1/2,0}(L)
       = \displaystyle\frac{1}{2}
         + \displaystyle\frac{1}{4}y_{0}(L)
         - \displaystyle\frac{1}{2}y_{\phi}(L)\,.
   \end{array}
  \label{eq:mag}
\end{equation}
Note that scaling dimensions $x^{\rm c,s}_{1/2,0}$ are not
the form (\ref{eq:dimm}).
This comes from the first order perturbation of the second term
($\cos\sqrt{2}\phi$ term) in equation~(\ref{eq:sg}).
Denoting $y_{\phi} = \pm y_{0}(1+w)$ where $w$ measures the distance
from the BKT transition point, we have for $y_{\phi}>0$
\begin{equation}
  \begin{array}{l}
     x^{\rm c}_{1/2,0}(L) = \displaystyle{1 \over 2}
          +\displaystyle{3 \over 4}
           y_{0}(L)\left( 1 + \displaystyle{2 \over 3}w \right) \\ \\
     x^{\rm s}_{1/2,0}(L) = \displaystyle{1 \over 2}
          -\displaystyle{1 \over 4}y_{0}(L)(1+2w)
  \end{array}
\end{equation}
and for $y_{\phi}<0$
\begin{equation}
  \begin{array}{l}
     x^{\rm c}_{1/2,0}(L) = \displaystyle{1 \over 2}
        -\displaystyle{1 \over 4}y_{0}(L)(1+2w)   \\ \\
     x^{\rm s}_{1/2,0}(L) = \displaystyle{1 \over 2}
        +\displaystyle{3 \over 4}y_{0}(L)
         \left( 1+\displaystyle{2 \over 3}w \right)\,.
  \end{array}
\end{equation}
On the other hand, from equation~(\ref{eq:dimn})
the scaling dimension of ${\cal{O}}_{0,\pm 2}$ is given by
\begin{equation}
  x_{0,\pm2}(L) = \frac{1}{2}-\frac{1}{4}y_{0}(L)
\label{eq:ele}
\end{equation}
from which we see that $x_{0,\pm2}$ and $x^{\rm c,s}_{1/2,0}$
(s for $y_{\phi}>0$ and c for $y_{\phi}<0$)
cross linearly at the transition point ($w=0$).

In order to identify the excitation with those of the sine-Gordon model
(\ref{eq:sg}), we can use the following symmetry.
The Hamiltonian with PBC is invariant under
the spin rotation around the $S^{z}$ axis,
the translation by three sites,
($\bi{S}_{j}\rightarrow \bi{S}_{j+3}$),
and space inversion
($\bi{S}_{j}\rightarrow \bi{S}_{L-j+1}$).
Corresponding eigenvalues are $M$, the wave number $q$, and $P=\pm 1$.
The space inversion in the sine-Gordon model are
\begin{equation}
  \phi\rightarrow -\phi~~~~~
  \theta\rightarrow \theta+\pi/\sqrt{2}~~~~~
  x\rightarrow -x\,.
\end{equation}
The magnetization $M$ is related to $n$ as $n=M_{\rm s}/3-M$.
The ``ground state'' energy $E_{g}$ is the lowest one with
$[M=M_{\rm s}/3, q=0, P=1]$.

In our model, the energy level corresponding to the operator
${\cal{O}}_{0,\pm2}$ is $E_0(M_{\rm s}/3 \pm 2,0,1)$,
where $E_0(M,q,P)$ is the lowest energy with $[M,q,P]$.
However, we cannot directly compare the energies with different $M$.
In the language of the spinless fermions, the difference in $M$
corresponds to the difference in the number of fermions, $N$.
Thus to compare the energies with different $M$,
we should use $E - \mu N$, where $\mu$ is the chemical potential
of the spinless fermions.
Since $\mu$ near $M_{\rm s}/3$ is expressed as
\begin{equation}
    \mu
    = {1 \over 4}
      \left\{E_0 \left( {M_{\rm s} \over 3}+2,0,1 \right)
           - E_0 \left( {M_{\rm s} \over 3}-2,0,1 \right) \right\}
    \label{eq:mu}
\end{equation}
the excitation energy corresponding to ${\cal O}_{0,2}$ is
\begin{eqnarray}
    \fl\Delta E_{0,2}
        &=& \left\{ E_0 \left( {M_{\rm s} \over 3}+2,0,1 \right)
           -  E_0 \left( {M_{\rm s} \over 3},0,1 \right)
           - 2\mu \right\} \nonumber \\
    \fl &=& {1 \over 2}
        \left\{ E_0 \left( {M_{\rm s} \over 3}+2,0,1 \right)
             +E_0 \left( {M_{\rm s} \over 3}-2,0,1 \right) \right\}
             -E_0 \left( {M_{\rm s} \over 3},0,1 \right)\,.
\label{eq:grand-canonical}
\end{eqnarray}
Just the same expression is obtained for ${\cal O}_{0,-2}$.
Equation~(\ref{eq:grand-canonical}) can be also obtained by use
of the Legendre transformation $E \to E - HM$.

The excitation energies corresponding to the operators
$\sqrt{2}\cos (\phi/\sqrt{2})$ and $\sqrt{2}\sin (\phi/\sqrt{2})$ are
obtained by the two lowest energies $\Delta E(M,P)$
with the twisted boundary condition as
\begin{equation}
  \begin{array}{l}
     \Delta E^{\rm c}_{1/2,0}
     = E^{\rm TBC} \displaystyle \left(\frac{M_{\rm s}}{3},1\right)
      -E\displaystyle\left(\frac{M_{\rm s}}{3},0,1\right)  \\ \\
     \Delta E^{\rm s}_{1/2,0}
     = E^{\rm TBC} \displaystyle\left(\frac{M_{\rm s}}{3},-1\right)
      -E\displaystyle\left(\frac{M_{\rm s}}{3},0,1\right)
  \end{array}
\end{equation}
where $E(M_{\rm s},0,1)$ is the lowest energy with PBC.
The excitation energies $\Delta E_{0,\pm2}$ and $\Delta E^{\rm c,s}_{1/2,0}$
(s for $y_{\phi}>0$ and c for $y_{\phi}<0$)
should cross linearly at the BKT transition point.
\begin{figure}[h]
   \begin{center}
      \scalebox{0.4}[0.4]{\includegraphics{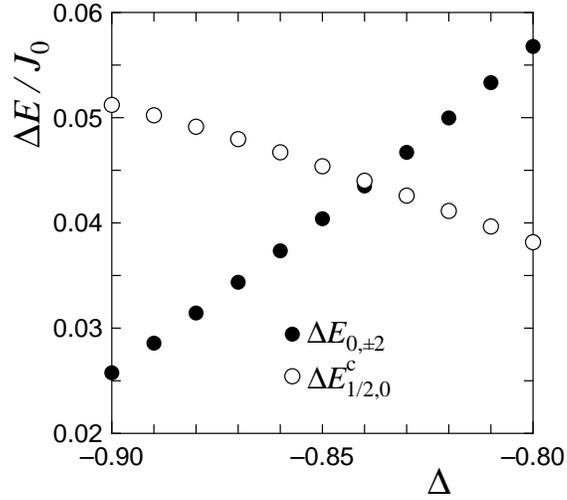}}
   \end{center}
   \caption{$\Delta E_{0,\pm 2}$ and $\Delta E^{\rm c}_{1/2,0}$
            for $L=18$ spins as functions of anisotropy parameter
            $\Delta$ when $t=-0.25$.
            From the crossing point
            we obtain $\Delta_{\rm c}(L=18)=-0.8389$.}
   \label{fig:level-cross}
\end{figure}

\begin{figure}[h]
   \begin{center}
      \scalebox{0.4}[0.4]{\includegraphics{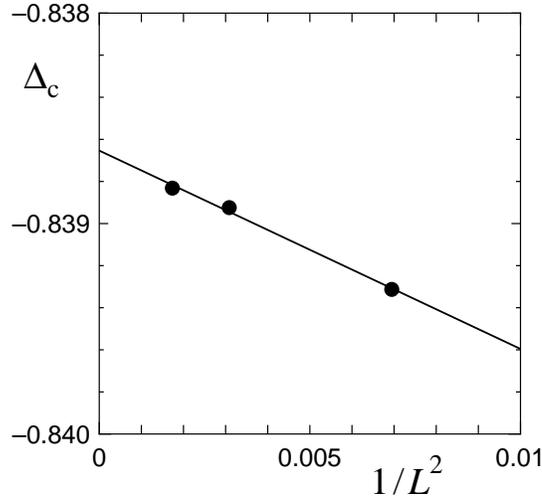}}
   \end{center}
   \caption{Extrapolation of $\Delta_{\rm c}$ to $L = \infty$
            when $t=-0.25$.
            We obtain $\Delta_{\rm c}=-0.839$.}
   \label{fig:delta-c}
\end{figure}

\begin{figure}[h]
   \begin{center}
      \scalebox{0.4}[0.4]{\includegraphics{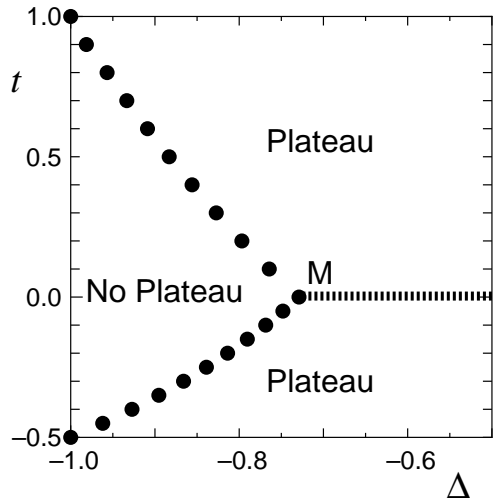}}
   \end{center}
   \caption{Phase diagram on the $\Delta-t$ plane.
      Closed circles are the BKT transition points determined
      from the numerical data as explained in the text.
      Thick dotted line denotes the Gaussian line.
      The multicritical point M corresponds to the point O
      of figure~\ref{fig:RG-flow}.}
   \label{fig:phase-diagram}
\end{figure}

Figure~\ref{fig:level-cross} shows the behaviour of $\Delta E_{0,\pm 2}$
and $\Delta E^{\rm c}_{1/2,0}$ for $L=18$
spins as functions of anisotropy parameter $\Delta$ when $t=-0.25$.
From the crossing point, we obtain $\Delta_{\rm c} = - 0.8389$ for $L=18$ spins.
The BKT transition point for the infinite system can be obtained
by extrapolating the $\Delta_{\rm c}$ data to $L = \infty$, as shown in
figure~\ref{fig:delta-c}.
Thus, we can obtain the phase diagram on the $\Delta-t$ plane
as shown in figure~\ref{fig:phase-diagram}.
The point M ($\Delta = -0.729$) is the multicritical point
where two BKT line meet together into the Gaussian line
(shown by thick dotted line)
on which the critical exponents vary continuously.
The $\Delta \le -1$ region is the ferromagnetic region.

Let us confirm the conformal anomaly $c=1$
which is related to the leading finite size correction of
the ``ground state'' energy with PBC as~\cite{BCN,Affleck}
\begin{equation}
  E_{\rm g}(L)
  = L \epsilon_{\rm g} - \frac{\pi v_{\rm s} c}{6L} + \cdots
  \label{eq:gs-correction}
\end{equation}
where $\epsilon_{g}$ the energy per one spin for the infinite size
system.
The spin wave velocity $v_{\rm s}$ can be obtained by
\begin{equation}
   v_{\rm s}
   = \lim_{L \to \infty}
     {L\Delta E (q = 2\pi / L)  \over 2\pi}
   \label{eq:vs-exc}
\end{equation}
where $\Delta E(q=2\pi/L)$ is the lowest excitation energy
having the wave number $q = 2\pi / L$ in the $M=M_{\rm s}/3$ space.
Thus we can check the value of $c$ by use of equations
(\ref{eq:gs-correction}) and (\ref{eq:vs-exc}).
We have found that $c=1$ is realized on the BKT line
within the error of a few percent.
For instance, in case of $(t,\Delta) = (0.5,-0.881)$ on the BKT line,
we obtain $v_{\rm s}c = 0.212 J_0$ through
equation~(\ref{eq:gs-correction}) and $v_{\rm s} = 0.217 J_0$
through equation~(\ref{eq:vs-exc}).
\begin{figure}[h]
   \begin{center}
      \scalebox{0.4}[0.4]{\includegraphics{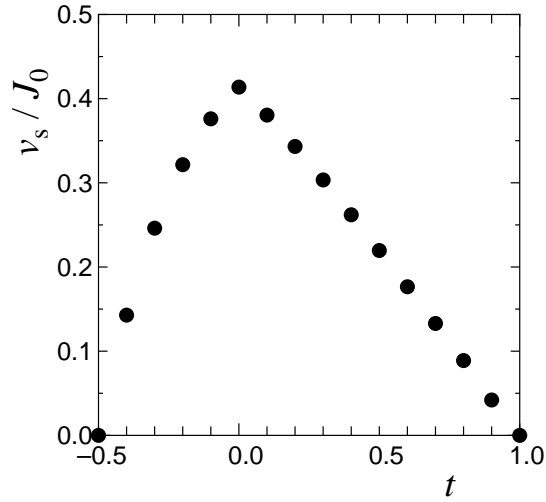}}
   \end{center}
   \caption{Spin wave velocity on the BKT line.}
   \label{fig:velocity}
\end{figure}
Figure~\ref{fig:velocity} shows the spin wave velocity
$v_{\rm s}$ on the BKT transition line.

\begin{figure}[h]
   \begin{center}
      \scalebox{0.4}[0.4]{\includegraphics{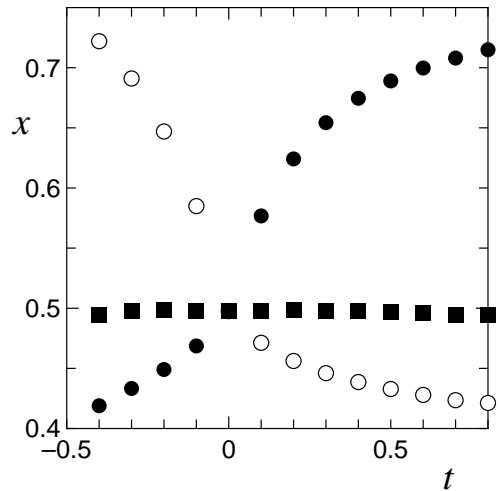}}
   \end{center}
   \caption{Scaling dimension on the BKT line.
            Closed circles are $x_{1/2,0}^{\rm c}(L)$,
            open circles $x_{1/2,0}^{\rm s}(L)$, 
            and closed squares the averaged scaling
            dimension~(\ref{eq:ave-x}).}
   \label{fig:ave-x}
\end{figure}

From equations~(\ref{eq:mag}) and (\ref{eq:ele}), we can eliminate the
leading logarithmic correction at the transition points ($w=0$)
using the following average
\begin{equation}
   \begin{array}{c}
     \displaystyle{3x_{1/2,0}^{\rm s}(L) + x_{1/2,0}^{\rm c}(L) \over 4}
     = \displaystyle{1 \over 2}~~~~~
          \mbox{for $y_{\phi}>0$}  \\ \\
     \displaystyle{3x_{1/2,0}^{\rm c}(L) + x_{1/2,0}^{\rm s}(L) \over 4}
     = \displaystyle{1 \over 2}~~~~~
          \mbox{for $y_{\phi}<0$}\,.
  \end{array}
  \label{eq:ave-x}
\end{equation}
This relation is appropriate to check our method of analyzing the numerical
data.
The averaged scaling dimension equation~(\ref{eq:ave-x}) on the BKT line
is shown in figure~\ref{fig:ave-x}.
We can see that the averaged
scaling dimension is very close to $1/2$,
which guarantees the consistency of our numerical method.


\section{Discussion}

We have obtained the phase diagram on the $\Delta-t$ plane as shown in
figure~\ref{fig:phase-diagram}.
Two BKT lines meet together into the Gaussian line at the multicritical
point M where $\Delta = \Delta_{\rm M} = -0.729$.
We can analytically predict the value of $\Delta_{\rm M}$
from $K=4$ with equations~(\ref{eq:K}) and (\ref{eq:A-C}).
We analytically obtain $\Delta_{\rm M} = -3\sqrt{3}\pi / 21 = -0.777$,
which shows fairly good agreement with the numerical value .
The analytically predicted value of the spin wave velocity at the
multicritical point is obtained by substituting
$\Delta_{\rm M} = -3\sqrt{3}\pi / 21$ into equation (\ref{eq:K}),
which results in $v_{\rm s} = 0.494 J_0$.

The slopes of the BKT lines for $t>0$ and $t<0$
near the multicritical point M are the same with each other.
This can be explained from the symmetry of
the bosonized Hamiltonian~(\ref{eq:sg}).
Hamiltonian~(\ref{eq:sg}) is invariant under the transformation $t \leftrightarrow -t$
and $\sqrt{2}\phi \leftrightarrow  \sqrt{2}\phi + \pi$.
As the multicritical point M is gone away, on the other hand,
the BKT lines on the upper and lower planes are asymmetric with each other,
as can be seen from figure~\ref{fig:phase-diagram}.
This is quite reasonable because the $t \leftrightarrow -t $ symmetry
does not hold in the original spin Hamiltonian~(\ref{eq:ham2}).
From the standpoint of the bosonized Hamiltonian,
this comes from the existence of higher order terms~\cite{Haldane}
$\cos(2\sqrt{2}\phi),\ \cos(4\sqrt{2}\phi), \cdots$,
of which coefficients are also proportional to the trimerization parameter $t$.
If these higher order terms are taken into account,
the symmetry of the bosonized Hamiltonian
under the transformation $t \leftrightarrow -t$ and
$\sqrt{2}\phi \leftrightarrow  \sqrt{2}\phi + \pi$ is lost,
which explains the asymmetry of the BKT lines.

\begin{figure}[h]
   \begin{center}
      \scalebox{0.4}[0.4]{\includegraphics{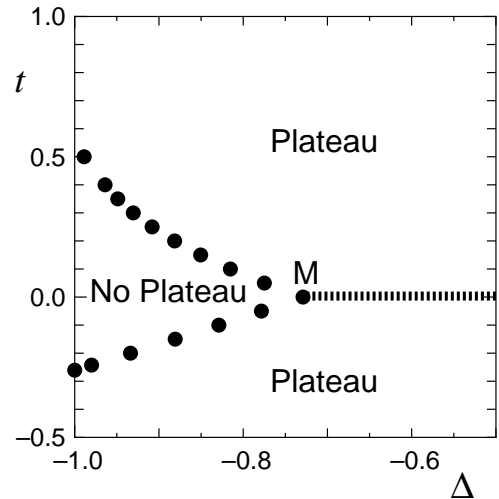}}
   \end{center}
   \caption{Phase diagram of the \lq\lq$xy$-trimerization model\rq\rq.}
   \label{fig:xy-phase-diagram}
\end{figure}

The mass-generating term $\cos(\sqrt{2}\phi)$
in the bosonized Hamiltonian~(\ref{eq:sg}) comes from the $J^x-J^x$
and $J^y-J^y$ couplings of the trimerization.
Strictly speaking, there exists another mass-generating term
\begin{equation}
   {2t\Delta \over \pi}
   \int \rmd x (\nabla \phi)^2 \cos(\sqrt{2}\phi)
   \label{eq:z-trimer}
\end{equation}
which comes from the $J^z-J^z$ coupling of the trimerization.
The effects of equation~(\ref{eq:z-trimer}) and the $\cos(\sqrt{2}\phi)$ term
in equation~(\ref{eq:sg}) are mutually competing when $\Delta < 0$.
In the $S=1/2$ ferromagnetic-antiferromagnetic alternating
chain~\cite{Okamoto-Nishino-Saika, Okamoto-F-AF},
this kind of competition brings about the transition
between the Haldane state and the large-$D$ state.
In our case, however, the term of equation~(\ref{eq:z-trimer}) only
works to reduce the coefficient of $\cos(\sqrt{2}\phi)$ in equation~(\ref{eq:sg}).
Most simple treatment may be the approximation
$(\nabla\phi)^2 \cos(\sqrt{2}\phi)
 \Rightarrow \langle (\nabla\phi)^2 \rangle \cos(\sqrt{2}\phi)$.
In fact, we obtain the phase diagram figure~\ref{fig:xy-phase-diagram}
for the \lq\lq $xy$-trimerization model\rq\rq \ in which
the trimerization exists only in the $J^x-J^x$ and $J^y-J^y$
couplings and not in the $J^z-J^z$ coupling.
We see that the trimerization effect is reduced in
figure~\ref{fig:phase-diagram} in comparison with
figure~\ref{fig:xy-phase-diagram}, because the no-plateau region
is wider in figure~\ref{fig:phase-diagram} than in
figure~\ref{fig:xy-phase-diagram}.
We note that the phase boundary of the ferromagnetic region
is no longer $\Delta = -1$, because the $SU(2)$ symmetry is
broken even at $\Delta =-1$ in the $xy$-trimerization model.
This situation is similar to the $S=1/2$ $XXZ$ chain under the
staggered magnetic field~\cite{Alkaraz-Malvetti}.
This phase boundary can be calculated from the instability
of the ferromagnetic state against the $M = M_{\rm s}-1$ spin wave
excitation, resulting in
\begin{equation}
    \Delta = -{1 + 2t + \sqrt{9-12t+12t^2} \over 4}\,.
\end{equation}

\ack
We would like to express our appreciation to K. Nomura and M. Oshikawa
for fruitful discussions.
A part of the numerical calculation was done by use of program package
TITPACK Ver.2 coded by H. Nishimori.
\end{document}